\documentstyle[12pt,psfig,epsfig]{l-aa}
\def\msun{M_{\odot}}
\def\ergsec{\hbox{erg s$^{-1}$}}

\def\degmark{^\circ}
\def \rsun {\ifmmode$R$_{\odot}\else R$_{\odot}$\fi}
\def \nh {N${\rm _H}$}
\def \hcm {\hbox {\ifmmode $ H atoms cm$^{-2}\else H atoms cm$^{-2}$\fi}}
\def \src {CAL\,87}
\def\approxgt{\mathrel{\hbox{\rlap{\lower.55ex \hbox {$\sim$}}
        \kern-.3em \raise.4ex \hbox{$>$}}}}
\def\approxlt{\mathrel{\hbox{\rlap{\lower.55ex \hbox {$\sim$}}
        \kern-.3em \raise.4ex \hbox{$<$}}}}
\newcommand {\rosat} {{ROSAT}}
\newcommand {\einstein} {{\it Einstein}}

\newcommand {\sax} {{\it BeppoSAX}}

\newcommand {\eg} {{e.g.}}

\newcommand {\Msun} {{ M$_{\odot}$}}

\newcommand {\rchisq} {$\chi_{{\rm \nu}} ^{2}$}

\begin{document}

\thesaurus{ (02.01.2; 08.02.1; 08.09.2; 08.23.1; 13.25.5)}

\title{A \sax\ observation of the super-soft source \src }

\author{A.N. Parmar\inst{1} \and P. Kahabka\inst{2} \and H.W.
Hartmann\inst{3} \and J. Heise\inst{3} \and D.D.E. Martin\inst{1} \and M.
Bavdaz\inst{1} \and T. Mineo\inst{4}}

\institute{Astrophysics Division, Space Science Department of ESA, 
ESTEC, P.O. Box 299, 2200 AG Noordwijk, The Netherlands
\and Astronomical Institute and Center for High Energy Astrophysics,
University of Amsterdam, Kruislaan 403, 1098 SJ Amsterdam, The
Netherlands
\and SRON Laboratory for Space Research, Sorbonnelaan 2, 
3584 CA Utrecht, The Netherlands
\and IFCAI/CNR, via U. Malfa 163, 90146 Palermo, Italy}

\date{Received ; accepted}
\offprints{A.N. Parmar: aparmar@astro.estec .esa.nl}
\maketitle

\begin{abstract}

We report on a \sax\ Concentrator Spectrometer
observation of the super-soft source (SSS) \src. The X-ray emission
in SSS is believed to arise from nuclear burning of accreted material
on the surface of a white dwarf (WD).
An absorbed blackbody spectral model gives a \rchisq\ of 1.18 
and a temperature of $42 \pm ^{13} _{11}$~eV. However, the derived 
luminosity and radius are greater than the Eddington limit and 
radius of a WD. 
Including an O~{\sc viii} edge at 0.871~keV gives a significantly
better fit (at $>$95\% confidence) and results in more realistic
values of the source luminosity and radius. We also fit 
WD atmosphere models to the \src\ spectrum. These also give
reasonable bolometric luminosities and radii in the ranges
2.7--4.8$ \times 10^{36}$~\ergsec\ and 8--20$\times 10^{7}$~cm, respectively.
These results support the view that the X-ray
emission from \src\ results from nuclear burning in the atmosphere of 
a WD. 

\end{abstract}


\section{Introduction}
\label{sec:introduction}

The \einstein\ observatory performed a survey of the Large Magellanic Cloud
(LMC) in which two sources with unusually soft spectra, CAL\,83 and \src\ were
detected (Long et al. 1981). These sources emit little or no radiation
at energies $\approxgt$1~keV and became known as ``super-soft'' sources
(SSS). Subsequent
\rosat\ observations have revealed approximately 30 similar sources
located in the Galaxy, the Magellanic Clouds, a globular cluster and M31
(see Kahabka 1995; Kahabka \& Tr\"umper 1996 for recent reviews). 
SSS are hard to detect in the galactic plane due to the effects of interstellar 
absorption. Absorbed black-body spectral models give typical 
temperatures of $\sim$40~eV and bolometric luminosities of 
$\sim$$10^{38}$~\ergsec.

SSS were originally interpreted as due to scattering 
from an accretion disk corona (e.g., Smale et al. 1988), or
accreting neutron stars radiating near or above the Eddington limit
(Greiner et al. 1991; Kylafis \& Xilouris 1993). Van den Heuvel et al. (1992)
proposed that these are systems undergoing
steady nuclear burning of hydrogen accreted onto the surface of a white 
dwarf (WD) with masses in the range 0.7--1.2\Msun. The mass transfer from
a main-sequence or sub-giant companions is unstable on a thermal 
time scale and for a narrow range of accretion rates, steady nuclear
burning can take place. Evolutionary scenarios for such systems are
discussed in Rappaport et al. (1994). It is unlikely that SSS
compose a homogeneous class and one way of probing the nature
of individual sources is by searching for the characteristic spectral
signatures of nuclear burning on a WD. This burning takes
place deep within the WD atmosphere at a large energy
dependent optical depth.
Photoelectric absorption by highly ionized metals in the atmosphere can
produce edges in the X-ray spectrum.
These effects have been modeled assuming Local Thermodynamic Equilibrium
(LTE) by Heise et al. (1994) and more recently extended to the
non-LTE (NLTE) case by Hartmann \& Heise (1997).  

\begin{figure*}
  \centerline{
       \hbox{\psfig{figure=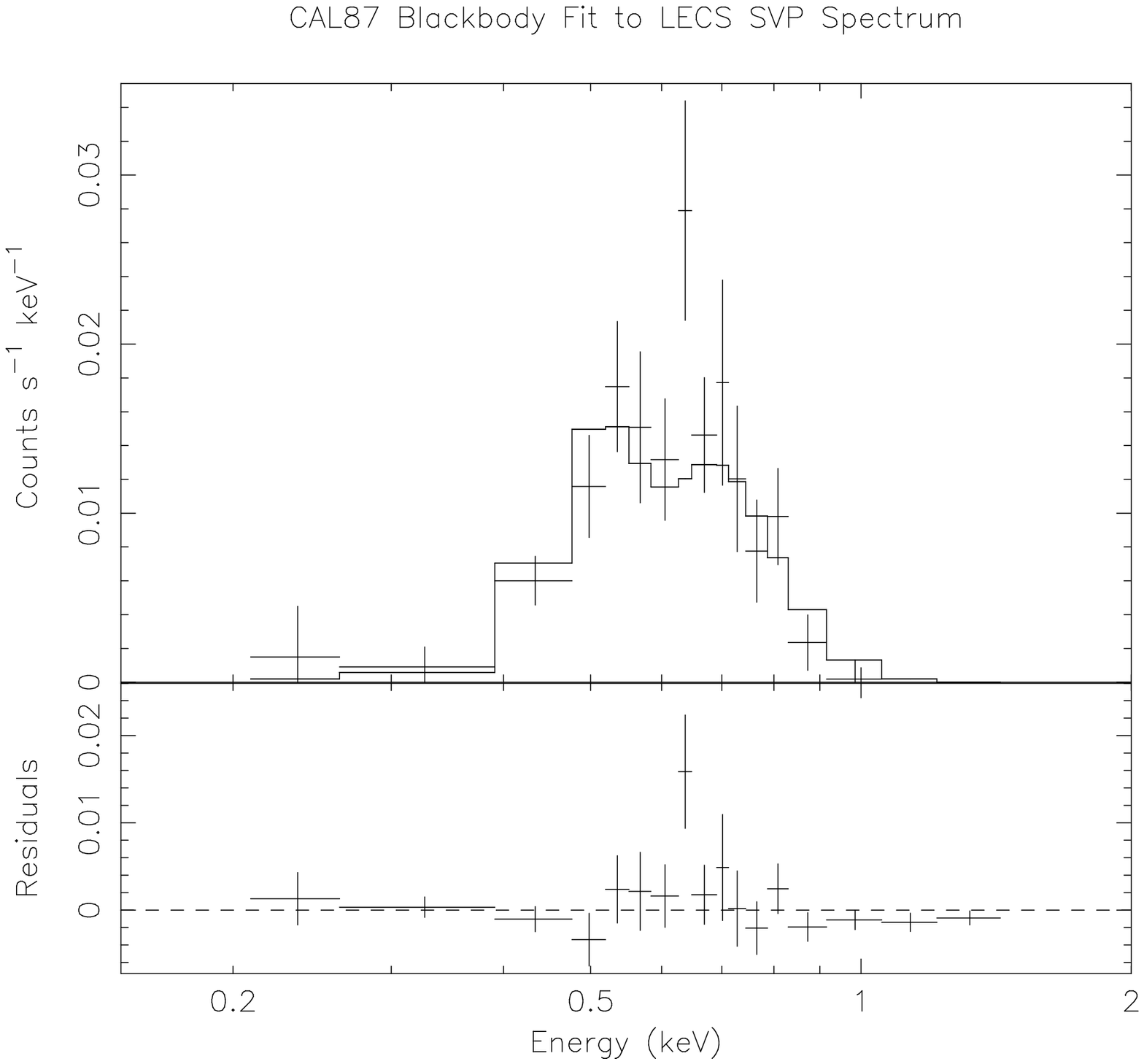,width=7.0cm,angle=-90,
             bbllx=50,bblly=50,bburx=575,bbury=650,clip}
             \psfig{figure=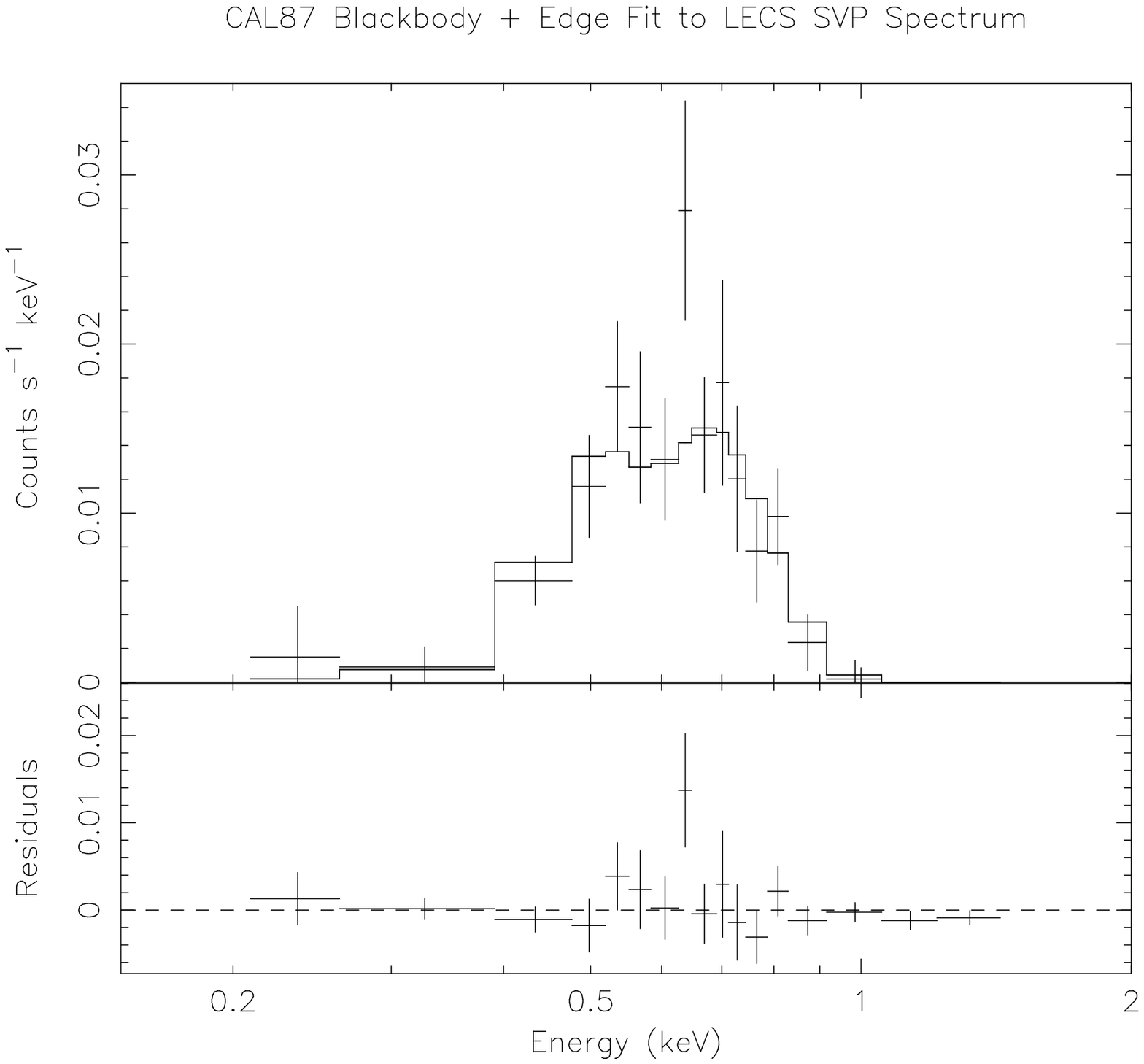,width=7.0cm,angle=-90,
             bbllx=50,bblly=50,bburx=575,bbury=650,clip}}}
  \caption{Absorbed blackbody (left panel) and blackbody with
a 0.871~keV absorption edge (right panel) model fits to the LECS \src\ 
spectrum}
  \label{fig:bb_fits}
\end{figure*}

\begin{figure*}
  \centerline{
       \hbox{\psfig{figure=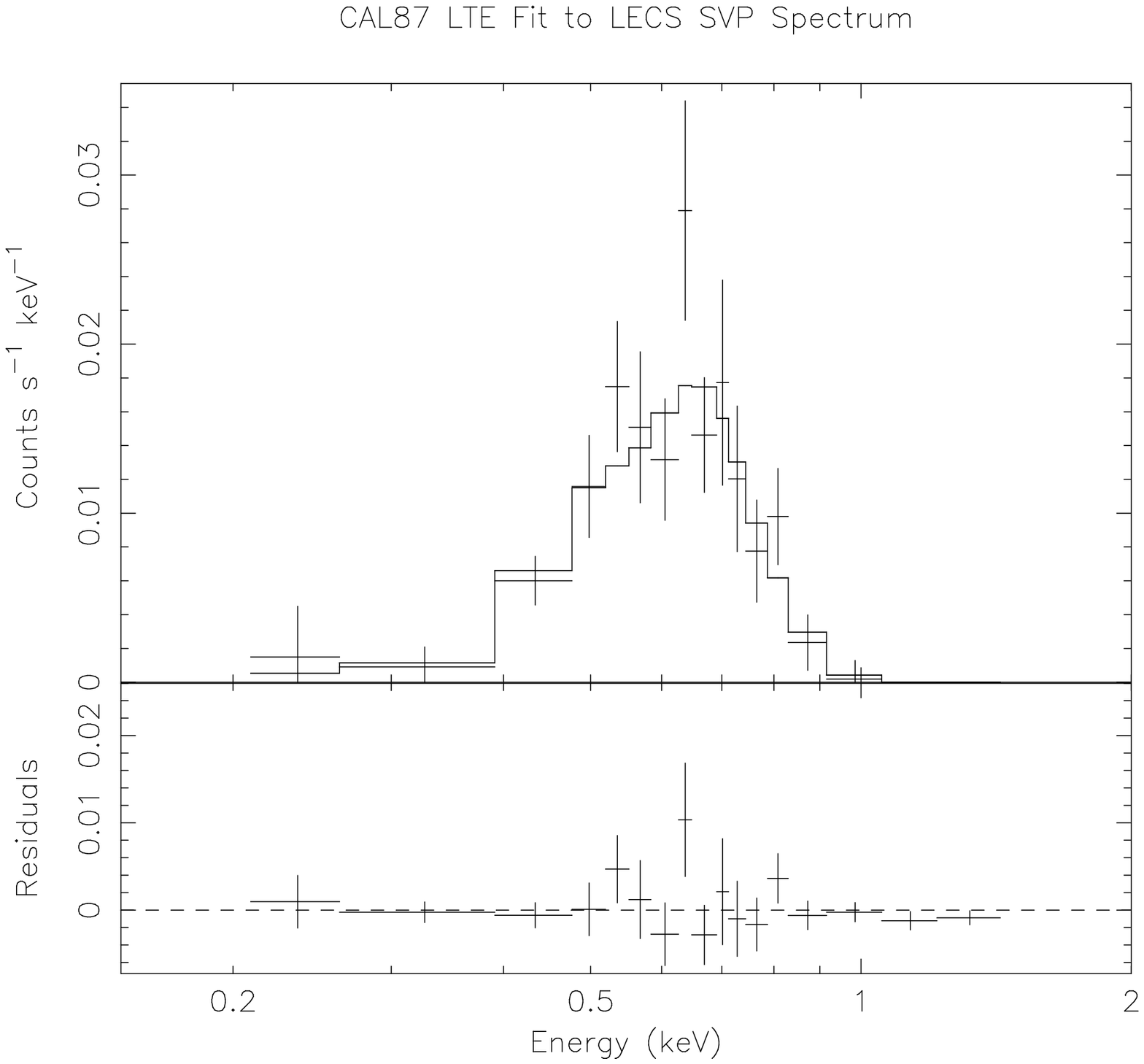,width=7.0cm,angle=-90,
             bbllx=50,bblly=50,bburx=575,bbury=650,clip}
             \psfig{figure=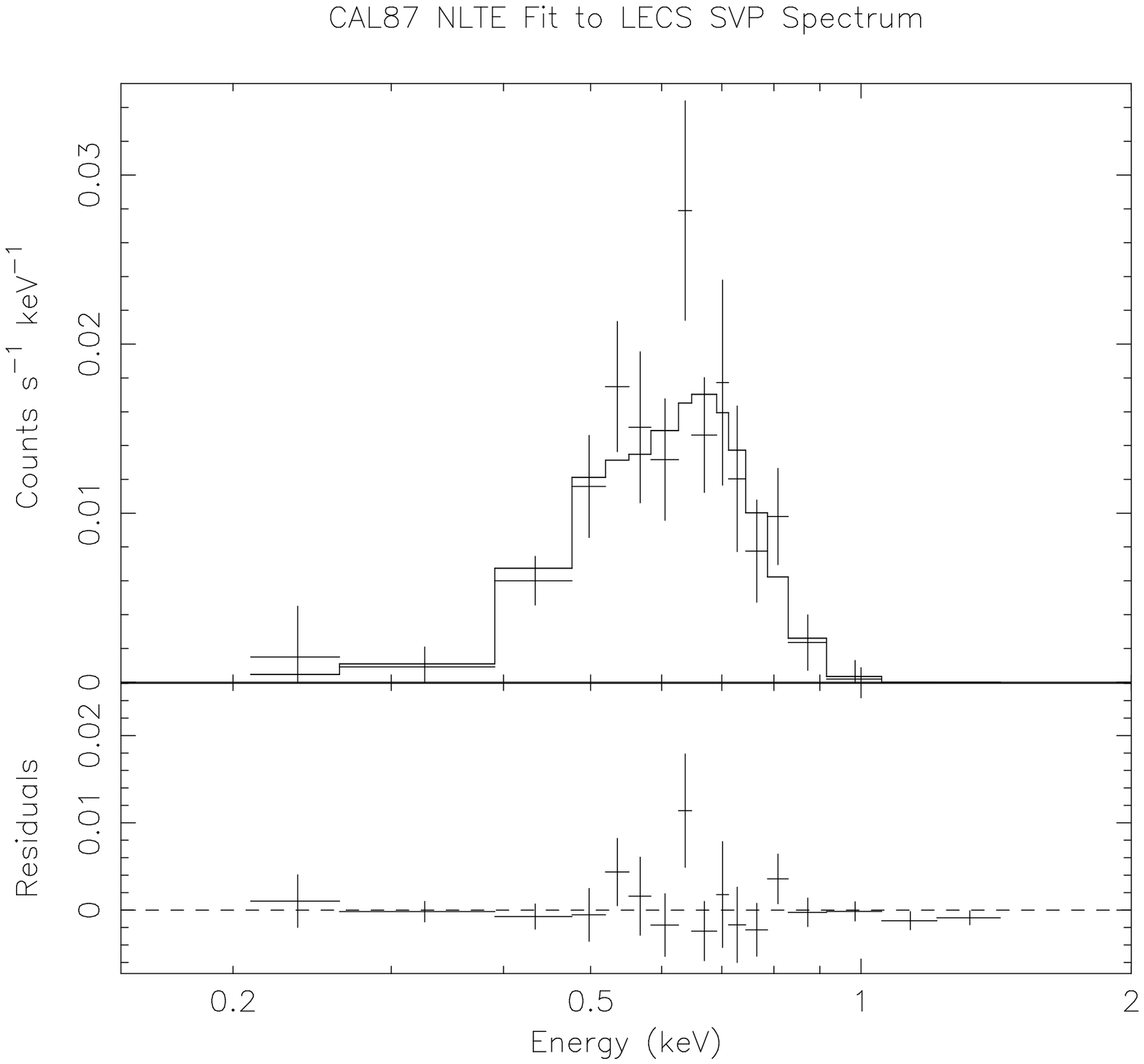,width=7.0cm,angle=-90,
             bbllx=50,bblly=50,bburx=575,bbury=650,clip}}}
  \caption{Absorbed LTE (left panel) and NLTE (right panel) model fits to 
  the LECS \src\ spectrum}
  \label{fig:model_fits}
\end{figure*}

\src\ exhibits both X-ray 
and optical eclipses with a period of 10.6~hrs (Callanan et al. 1989; 
Cowley et al. 1990; Schmidkte et al. 1993; Kahabka et al. 1994), indicating 
an orbital inclination of $>$$70\degmark$. The optical lightcurve shows a
deep asymmetric primary minimum with a shallow, variable, secondary
minimum while the X-ray eclipse is narrower and shallow.
The optical modulation may be due to obscuration by a structured accretion
disk (Schandl et al. 1997).  
Fitting an absorbed blackbody model to a 37 ksec exposure \rosat\ Position
Sensitive Proportional Counter (PSPC; 0.1--2.5~keV; Tr\"umper 1983) 
\src\ spectrum gives a best-fit
temperature, T, of 35~eV and an equivalent hydrogen column density, \nh,
of $1 \times 10^{22}$~\hcm\ (Hartmann \& Heise 1997). 
\src\ has been a persistent X-ray source since its discovery in 1980.

\section{Observations}
\label{sec:observations}

The scientific payload of the \sax\ X-ray Astronomy Satellite 
(Boella et al. 1997a) comprises 
four coaligned Narrow Field Instruments, or NFI, including the Low Energy
and Medium Energy Concentrator Spectrometers (LECS and MECS).
The LECS is an 
imaging gas scintillation proportional counter sensitive in 
the energy range 0.1--10.0~keV with a circular field of view of 37$'$ 
diameter (Parmar et al. 1997). The background counting rate is
$9.7 \times 10^{-5}$~arcmin$^{-2}$~s$^{-1}$ in the energy range 
0.1--10.0~keV.
The LECS energy
resolution is a factor $\sim$2.4 better than that of the \rosat\ PSPC,
while the effective area is between a factor $\sim$6 and 2 lower at 0.28 and
1.5~keV, respectively.   
\src\ was observed by the LECS as part of the Science Verification Phase 
between 1996 October 27 02:14 to October 28 23:42~UTC. Due to the failure of 
a ground segment link and an instrument anomaly, data between 
October 28 04:54 and 10:55~UTC were lost.
Good data were selected from intervals when the minimum elevation angle
above the Earth's limb was $>$4$\degmark$ and when the instrument's settings
were nominal using the SAXLEDAS 1.4.0 data analysis package.  
Since the LECS was only operated 
during satellite night-time, this gave a total on-source exposure of
39~ksec. The MECS is
sensitive in the energy range 1.5--10~keV, with energy and angular 
resolutions similar to the LECS (Boella et al. 1997b). The MECS
observed \src\ for a total exposure of 120~ksec, but did not
detect the source.

Examination of the LECS image shows a source at a position consistent
with (32$''$ distant) that of \src.
A spectrum was extracted centered on the source centroid using a
radius of 8$'$. This radius was chosen to include 95\% of the 0.28~keV
photons.
The spectrum was rebinned to have $>$20
counts in each bin to allow the use of the $\chi^2$ statistic. The XSPEC
9.01 package (Arnaud 1996)
was used for spectral analysis together with the response matrix 
from the 1996 December 31 release.
Background subtraction was
performed using a standard blank field with a 46~ksec exposure. 
The \src\ count rate above background was 
$0.0076 \pm 0.0011$~s$^{-1}$. Examination of the extracted spectrum
shows that the source was only detected in a narrow energy range (see
Figs.~\ref{fig:bb_fits} and ~\ref{fig:model_fits}) and only the 17 
rebinned channels 
corresponding to energies between 0.2 and 1.5~keV were used for spectral 
fitting. 
\subsection {Spectral fits}
\label{subsec:spectrum}
In order to compare the LECS spectrum with those obtained from previous
observations, we first fit an absorbed blackbody spectral model to the data
(Fig.~\ref{fig:bb_fits}).
The photoelectric absorption coefficients of Morisson \& McCammon (1983) 
together with the solar abundances of Anders \& Grevesse (1989) were used.
An acceptable fit is obtained with \rchisq\ of 1.18 for 14 degrees
of freedom (dof). The best-fit parameters are given in Table~\ref{tab:bb_fits}.
A distance of 50~kpc is assumed in order to
derive the WD radius, R, and luminosity, L, and
all uncertainties are quoted at 68\% confidence.
The spectrum shows evidence for an abrupt cutoff $\approxgt$0.8~keV
and so an O~{\sc viii} edge with absorption depth $\tau$ at 
a fixed energy of 0.871~keV,
was added to the model. This edge is
the dominant spectral feature at energies $\approxgt$0.8~keV in 
many WD model atmosphere calculations (\eg\ Heise et al. 1994; White et al.
1995).
Including the edge gives a higher best-fit
temperature and improves the fit quality, 
resulting in a \rchisq\ of 0.93 (Table~\ref{tab:bb_fits}).
The value of the F statistic of 4.76 indicates that the addition of the 
edge is significant at $>$95\% confidence. If the edge energy
is allowed to vary, then the best-fit value of 
$0.84 \pm {0.04}$~keV is consistent with an O~{\sc viii} edge.

Heise et al. (1994) show that optically thick plasmas
in the temperature range $10^5 - 10^6$~k
are more efficient soft X-ray (0.1--1 keV) emitters than
blackbodies, assuming plane parallel hydrostatic model atmospheres
in which LTE determines the degree of ionization.
This conclusion has been extended to the NLTE case by 
Hartmann \& Heise (1997) for both solar and LMC abundances.
These models include free-bound opacity sources for all
relevant ions, but are still only first order approximations
since line blanketing has not been taken into account.
In addition, close to the Eddington limit the assumptions of 
hydrostatic equilibrium and plane parallel atmospheres are no longer
valid.

The above LTE and NLTE models were fit to the LECS \src\ spectrum.
For the LTE case, we assume an LMC abundance of 0.25 times the solar
value and a local gravity of $\log(g) = 9$, appropriate
to WDs with mass $\geq 0.6 \msun$. The fit results are however
insensitive to abundance and adopting solar abundance gives almost
identical results.
Models with $\log(g) \leq 8.5$ cannot be made hot enough in
hydrostatic equilibrium (due to the Eddington limit) to fit the spectrum.
We note that models with $\log(g) >> 9$ cannot be excluded since they
can fit the spectrum at higher effective temperatures
and lower source radii.

Assuming a power-law spectrum with a 
photon index of 2.09 (i.e. similar to that of the Crab Nebula) and a distance
of 50~kpc, the 99\% confidence upper-limit to any 2.0--10.0~keV
emission from \src\ obtained using MECS data is $1.3 \times 10^{34}$~\ergsec.

\begin{table}
\caption[]{\src\ blackbody, LTE (Heise et al. 1994) and NLTE (Hartmann \& 
Heise 1997) WD model atmosphere spectral fit parameters}
\begin{flushleft}
\begin{tabular}{lll}
\hline\noalign{\smallskip}
& Blackbody & Blackbody with\\
&           & 0.871~keV Edge  \\
\noalign {\smallskip}
\hline\noalign {\smallskip}
      T (eV)                 & $42\pm ^{13} _{11}$ & $59\pm ^{27} _{17}$ \\
      \nh\ ($10^{22}$ cm$^{-2}$) & $1.00 \pm ^{0.05} _{0.11}$ & 
                                          $0.53 \pm ^{0.58} _{0.02}$ \\
      $\tau $                 & \dots & $>$13 \\
      R (cm)                  & $2 \times 10^{9}$ - $7 \times 10^{12}$ & 
                                $ 7 \times 10^7$ - $6 \times 10^{10}$ \\ 
%      L  (erg s$^{-1}$)       & $4 \times 10^{38}$ - $6 \times 10^{44}$ & 
%                                $4 \times 10^{36}$ - $1.5 \times 10^{41}$  \\
      L  (10$^{36}$ erg s$^{-1}$)& 400 - $6 \times 10^{8}$ & 
                                4 - $1.5 \times 10^{5}$  \\
      \rchisq\                & 1.18 & 0.93 \\
      dof                     & 14 & 13 \\
\noalign {\smallskip}
\hline
\noalign {\smallskip}
& LTE & NLTE \\
\noalign {\smallskip}
\hline\noalign {\smallskip}
      T (eV)  & $74.4\pm 1.7$   & $57.3 \pm ^{1.9} _{2.4}$  \\ 
      \nh\ ($10^{22}$ cm$^{-2}$) & 
                   $0.18 \pm ^{0.12} _{0.06}$ & $0.19 \pm ^{0.17} _{0.07}$ \\
      R (cm)   & $(9.1 \pm 1.2) \times 10^7$   
      & $(1.67 \pm ^{0.35} _{0.25}) \times 10^8$ \\
%      L  (erg s$^{-1}$) & $(3.30 \pm 0.54) \times 10^{36}$
%      & ($3.89 \pm ^{0.89} _{0.69})\times 10^{36}$ \\ 
      L  (10$^{36}$ erg s$^{-1}$) & $3.30 \pm 0.54$ 
      & $3.89 \pm ^{0.89} _{0.69}$ \\ 
      \rchisq\      & 0.76 & 0.76 \\
      dof           & 14  & 14 \\
\noalign {\smallskip}
\hline
\end{tabular}
\end{flushleft}
\label{tab:bb_fits}
\end{table}

\section {Discussion}
\label{subsec:discussion}
The LECS spectrum of \src\ can be represented by all four types of 
trial models and it is clear that a LECS spectrum with significantly 
greater exposure is required to meaningfully discriminate between 
these models based on fit quality alone.  
There are differences in the best-fit values of T determined using 
the different models (see Table~\ref{tab:bb_fits}), with the blackbody
fit giving the lowest T (and hence the largest source
radius and luminosity) and the LTE fit the highest. The relatively large
uncertainties in the best-fit parameters means that the luminosity and
size of the X-ray emitting region are poorly constrained.
The values of \rchisq\ favor the interpretation of the spectrum in terms 
of a model atmosphere fit
with the LTE and NLTE fits both giving \rchisq\ values of 0.76. 
This should be compared with the blackbody fit which 
gives a \rchisq\ of 1.18. Both WD model atmosphere fits imply similar
values of \nh, while the temperature derived from the NLTE fit is
significantly cooler ($57.3 \pm _{2.4} ^{1.9}$~eV) than that derived assuming
LTE ($74.4 \pm 1.7$~eV). 
The best-fit blackbody T derived here of $42 \pm ^{13} _{11}$~eV
is slightly higher, but consistent with, those derived using the \rosat\
PSPC of $31 \pm ^{11}_{10}$~eV and $34 \pm ^{8}_{10}$~eV (Kahabka et al.
1994). 

The bolometric luminosity
implied by the blackbody interpretation of $>$$4 \times 10^{38}$~\ergsec\ 
is higher than the Eddington luminosity for a $1\msun$ object of
$1.3 \times 10^{38}$~\ergsec. In addition, the required blackbody radius
of $>$$2 \times 10^{9}$~cm is significantly larger than the expected 
WD radius ($8.7 \times 10^8$~cm for a $0.6\msun$ WD; Nauenberg 1972).
In contrast, the fits using WD 
atmosphere models imply a lower luminosity, radius and temperature of 
2.7--4.8$ \times 10^{36}$~\ergsec, 8--20$ \times 10^{7}$~cm
and 55--76~eV, respectively.
The WD mass can be constrained assuming that \src\ is on the
stability line (see Iben 1992, Fig.~2). The above temperature
range corresponds to a WD of mass $\sim$1.2$\msun$ which has a 
luminosity of 4--8$ \times 10^{37}$~\ergsec\ while undergoing
steady nuclear burning 
(see also Iben \& Tutukov 1996). This is at least a factor 8 
greater than the luminosity derived above. 
Since \src\ has
an orbital inclination of $>$$70\degmark$, it is possible 
that part of the emitting region is obscured, perhaps by the accretion
disk.
The LECS spectrum of \src\ is therefore consistent with
the assumption of a hot WD atmosphere heated by nuclear burning, but 
formally does not prove such an assumption.

\begin{acknowledgements}
We thank the \sax\ Mission Director R.C. Butler and P. Giommi
and F. Fiore for help with these
observations. The referee, S. Rappaport, is thanked for helpful
comments.
PK is a Human Capital and Mobility Fellow.
The \sax\ satellite is a joint Italian and Dutch programme.
IFCAI are supported by the Italian Space Agency (ASI) in the
framework on the \sax\ mission.
\end{acknowledgements}

\end{document}